\documentstyle[11pt,fleqn]{article}
\title{Possible Pressure Effect for Superconductors}
\author{A. Kwang-Hua Chu \thanks{Corresponding address after 2007-Aug-31 for the author :
P.O. Box 39, Tou-Di-Ban, Urumqi 830000, PR China.}}
\date{24, Lane 260, Section 1, Muja Road, Taipei, Taiwan 11646, ROC}
\begin{document}           
\maketitle
\begin{abstract} We make an estimate of the possible range of
$\Delta T_c$ induced by high-pressure effects in post-metallic
superconductors by using the theory of {\it extended
irreversible/reversible thermodynamics} and Pippard's length
scale. The relationship between the increment of the
superconducting temperature and the increase of the pressure is
parabolic. \newline

\end{abstract}
\doublerulesep=6mm        
\baselineskip=6mm  %
\oddsidemargin=-1mm  
\bibliographystyle{plain}  
\section{Introduction}  
The investigation of electron motions in polar crystal began 60
years ago \cite{Froehlich:1939}. Their influence in the phenomenon
of pressure-enhanced superconductivity has attracted the attention
of many researchers \cite{Boughton:Pressure}. The work reported in
[2] was limited to the study of low-temperature superconductivity
in periodic systems. Recently, related studies of High-Temperature
Superconductivity (HTS) in the cuprates \cite{Galasso:Perovskite}
and MgB$_2$ [4-5] became the focus of attention. \newline It was
the unusual high pressure effect on $T_c$ observed in (La,
Ba$)_2$CuO$_4$ that first signaled the significance of cuprates in
the search for high $T_c$ and led to the discovery of
YBa$_2$Cu$_3$O$_7$ and related compounds.  Meanwhile, as reported
in Ref. [5] for MgB$_2$, the superconducting transition
temperature $T_c \sim M^{\alpha_B}$, with $\alpha_B =0.26$ ($M$ is
the isotope mass) confirms the isotope relation for the important
role that phonons play in traditional superconductors (which the
BCS theory could be applied for). In addition, band structure
calculations indicate a rather isotropic electrical transport
instead of the very layered appearance of honeycombed boron and
hexagonal magnesium networks in the material [6]. These facts
support our present approach which will be described below. For
simplicity, we will not consider the effect due to
pressure-induced-charge-concentration in this study since it is
beyond our present approach and interest.
\newline Theoretically, from the theory of Quantum Mechanics
\cite{QM:Landau}, which in general treats very-small-scale
particles/waves, however, we have the pressure defined by $p=F/A$
with $F=(\partial H/\partial \lambda)_{nn}$ $=\partial
E_n/\partial \lambda$, where $\lambda$ is a parameter on which the
effective Hamiltonian $H$ (and therefore the energy eigenvalues
$E_n$) depends. Here, $H$ comes from the pressure effect, $A$ is
the area which is normally calculated with an artificial
surface-cut across the characteristic domain,, $F$ is the force
acting upon $A$; $n=0,1,2,\cdots$. The kinetic pressure from the
(mostly spherical) particles acting upon $A$ is usually presumed
equal to the hydrostatic pressure or the average of the trace from
the stress tensor (on $A$) in thermodynamic equilibrium so that we
can interpret the measurements easily and conveniently [8-9]. The
trouble along this thinking is : how to find this $H$ directly so
that we could interpret the pressure-effect easily?
\newline Griessen \cite{Griessen:HP}  discussed a number of models
which tried to explain the effect of pressure on the equilibrium
phase transition of superconducting properties [10-13]. A
theoretical approach to the pressure-effects for cuprates was
recently discussed in \cite{Angilella:Hole} using a BCS-type mean
field approach. Unfortunately most of the detailed structures of
the high-pressure phases are not yet known. \newline  To the best
of our knowledge, the role of the pressure-gradient, which may
drive electrons and/or phonons into a flow before the final
equilibrium is reached, in the study of kinetic or non-stationary
effects in superconductors is seldom mentioned [6,15]. In this
{\it Letter}, we shall use the idea of {\it Extended
Reversible/Irreversible Thermodynamics} (ER/IT)\cite{Mueller:EIT}
to estimate the possible range of $\Delta T_c$ from the
pressure-gradient resulting from the effect of uniaxial stress on
$T_c$ in superconductors. {\it In the frame of ER/IT, thermodynamic
functions or potentials can be functions of the gradient of
thermodynamic state variables as well as of thermodynamic state
variables only }[8-9]. \newline Once the external pressure is
imposed upon the sample, due to the highly anisotropic and
heterogeneous feature of the material, there will be a net
pressure- or density-gradient acting upon electrons (gases) inside
a presumed very-small slender domain (considering the pressure
from both longer sides of it) which may be bounded by near-by
phonons. The unbalanced pressures imposed upon the inlet and
outlet of a nano-channel (along which the electron gases pass
through) in our consideration can thus give the pressure- or
density-gradient and then drive the electron-gas flow. Our
interests here are those stationary states or steady electron-gas
flows so that we could consider the nearly equilibrium properties.
\newline We assume the BCS \cite{BCS:LTS} theory could be extended
to the situations after the samples are imposed upon high
pressures (they are already superconducting for whatever mechanism
which we have no interests in investigating here) and in certain
sense still valid here \cite{Griessen:HP} for some periodic
microdomains and the pressures or uniaxial stresses imposed on the
samples can influence the passing through of the electron-pairs so
that the superconductivity shifts with $T_c$. The overall effects
of external stresses on phonons are presumed to be completely
transmitted to those electron-pairs. Thus, we only need to
consider the influences to electron-pairs from those imposed
pressure or stress fields within their limits [17-18]. We must
also assume, however, that there were no created micro- or
nano-cracks inside during the imposing of the high pressures in
the cuprates for previous measurements so that our approach
described below could be well applied.   
\section{Formulation}  
In a strict analysis of transport problems it is seldom possible
to deal solely with average properties such as the mean velocity
and the mean energy, and it is necessary to determine the
distribution of the particles both as regards position and their
velocities. The fundamental equation determining the distribution
function is an integro-differential equation known as the
Boltzmann equation [17]
\begin{displaymath}
 \frac{\partial f}{\partial t}+\frac{{\bf F}}{m}\cdot \frac{\partial
 f}{\partial {\bf v}}+{\bf v}\cdot \frac{\partial f}{\partial {\bf x}} =G-L,
\end{displaymath}
where $G$, $L$ means gain and loss of the number of electrons due
to collisions; ${\bf F}$ is the external field force, $m$ is the
mass. If the electron gas is dilute enough, then we can neglect
the collisional integral (r.h.s. term of above equation).
Furthermore, by considering only the stationary state or final
equilibrium state, we can omit the first term in the
left-hand-side of the above equation. We also assume that the
contribution from ${\bf F}$ is much weaker than the
pressure-induced (into the velocity) term in the above equation.
The complex boundary-interactions, e.g. electron-phonon scattering
along the boundary, are also excluded.
\newline Using an extension of BCS theory, a general expression
has been derived  \cite{Kresin:BCS} $ T_c = 0.25 \tilde{\Omega}
(e^{2/\eta} -1 )^{-1/2}$, where $\tilde{\Omega} =\langle \Omega^2
{\rangle}^{1/2}$ stands for the characteristic phonon frequencies
and sets the energy scale ($\sim$ the Debye temperature for
certain cases), $\eta$ is the strength of the electron-phonon
coupling. The weak-coupling BCS formula \cite{BCS:LTS} gives $ T_c
=\omega_D \exp \{ 1/[N(E_F) V] \}$, where $V$ is the attractive
interaction due to the exchange of phonons, $N(E_F)$ is the
Density of States (DOS) at the Fermi energy $E_F$, $\omega_D$ is
the energy scale. The direct relations between $\Delta T_c$ and
$\Delta p$ \cite{Heisenberg:1949}, however, cannot be easily
obtained up to now [6,14].
\newline
One of the crucial parameters for our approach is related to the
density-gradient driven speed (say, $\bar{v}$)[9,21] or flux of
electrons in the microdomain when the phonons are stiffened by the
imposing pressure in the prescribed direction. This flux will then
be linked to the resistivity, and thus finally $T_c$ as the
equilibrium is reached. There is, however, one fundamental length
: the extended Pippard's coherence distance $\xi_0$, which is
associated by means of the uncertainty principle with the energy
$k T_c$ \cite{deGennes:1966},
\begin{equation}
 \frac{\hbar v_F}{\xi_0} \sim k T_c.
\end{equation}
$\hbar=1.0546 \times 10^{-34}$ J s, $k=1.38 \times 10^{-23}$ J
K$^{-1}$. This length scale can give us clues about the
pressure-induced correlation length which is the distance beyond
which the momenta are essentially uncorrelated.  \newline We now
let $T_c$ be a function of $d p/dx$ as well as $p$, which is valid
by the assumptions of {\it Extended Reversible/Irreversible
Thermodynamics} [8-9,21] that the final state is not far from the
statistical equilibrium. $x$ is linked to the effective distance
for the pressure imposed and within this range the electron-gas
flow is weakly compressible \cite{Laughlin:QF} and
fully-developed.  \newline Furthermore, as mentioned above, the
BCS theory is presumed also valid over the region we considered
(already post-metallic) after the imposing high-pressure. Then, by
considering the equilibrium limit of the evolutional
electron(pair)-gas motion (characterized by $\bar{v}$), and taking
the limit of $\bar{v} \sim v_F$, besides, as $\bar{v}=K_a c$ ($c$
: the sound speed $\equiv \sqrt{(dp/d\rho)|_s}=\sqrt{dp/dx \cdot
dx/d\rho}$; the latter relation is under the ER/IT formulations)
or $\bar{v}$ $\sim K_0 |(dp/dx)|^{0.5}$ (to certain limit of the
flux [9,21]), so with (1), $T_c$ may be linked to $|dp/dx|$ for
some situations \cite{Locquet:LSCO}. This consideration can be understood
that there is certain resonance existing between the propagating phase-speed of the interface
of electron pairs and the phonons. This resonance is induced by the possible localization due to the dynamical environment near-by. From equation (1) with the
effective length scale being $O(\xi_0)$ and by considering the
data re-arrangement, i.e. $\Delta T_c$ (K) vs. $\Delta p$ (GPa),
we could obtain the net increase of $T_c$ (K) due to the imposed
pressure p (GPa) by
\begin{equation}
 \Delta T_c = \kappa (\Delta p)^{0.5},
\end{equation}   
where $\kappa$ is strongly dependent on $dx/d \rho$ for unit width
($\rho$ is the density; we take the average of long-range
correlations). This expression might be extended and thought of an
possible limit for those anisotropic cases which are common in HTS
cuprates or other type of superconductors [24-27]. Note that once
the moving particles or pairs are composed of holes, the sign convention for
the density $\rho$ and the (local) coordinate $x$ (as the coordinate system for
quasi-one-dimensional electron motion has been prescribed and fixed) should be thus
changed as the hole motion is opposite to that of the electron. Meanwhile, considering
the definition of the concentration and the pressure of particles (pairs), the sign
convention will be reversed once the electron is replaced by the hole. It means
if $dp/dx$ is positive for the motion of electrons then $dp/dx$ is negative for that
of holes. The subsequent result is that $\Delta T_c$ could be either positive or negative! 
\section{Results and Discussions}  
We obtain an {\it ad hoc} estimate of the limit of $\Delta T_c$
(K) vs. $\Delta p$ (GPa) from equation (2) for some
superconductors which could be possibly extended to HST if
they are already post-metallic (thus the BCS theory then applies
for [6]) under very-high pressures. This approach might be
universal for similar superconductors and independent of the
experimental procedures (except the selection or tuning of
$\kappa$ [9,27]). This is because, at least in part, $T_c$ is
related to the equilibrium phase transition which in general has
no close link to the flow-history, and partly, $|dp/dx|$ may
change the mean free path ($\lambda$) of the electrons but the
latter is independent of $T_c$ \cite{Anderson:MFP} in most cases.
The presence of inequivalent layers may lead to more complicated
$\Delta T_c$ v.s. $\Delta p$ curves.
\newline The interesting thing is : our result has the similar
trend with that of [6] and \cite{Scholtz:1992} obtained from the
series of measurements at increasing and decreasing pressure, for
different types of superconductors, respectively. In fact, the
granular limit of superconductivity under very high pressure could
be reached once the perovskite structure of HTS suffers
considerable changes [24]. If the resonance induced by the possible localization
is relaxed due to a delocalization then the power (of $\Delta p$) shown in the equation of (2) will not be $0.5$!
 Our approach will not be valid,
however, once the dissipation produced by the pressure-driven flow
of electron gas is too large so that the basic assumptions of
the{\it Extended Reversible/Irreversible Thermodynamics} [8-9]
being violated. In that case, we cannot predict in which way $T_c$
starts to decrease (or increase) even the pressure still
increases. Perhaps, the approach of pressure-induced charge
transfer (effect) [14,30] might help us understand the HTS
behavior for this kind of large-dissipation flow \cite{Chu:2000}
of electron gases. We shall investigate more complicated problems [32]
in the future.

\subsubsection*{Acknowledgements}
The main approach of this research (together with Ref. [27]) was
initiated when the author stayed in Muja around the end of 1998.
\end{document}